\documentclass[aps,prl,twocolumn,showpacs,superscriptaddress,nofootinbib,10pt]{revtex4-2}

\usepackage{acronym}
\usepackage{amsfonts}
\usepackage{amsmath}
\usepackage{amssymb}
\usepackage{bm}
\usepackage{cases}
\usepackage{color}
\usepackage{comment}
\usepackage[dvipsnames]{xcolor}
\usepackage{enumerate}
\usepackage{epsfig}
\usepackage{etoolbox}
\usepackage{fancyref}
\usepackage{fancyhdr}
\usepackage[T1]{fontenc} 
\usepackage{graphicx}
\usepackage{hyperref}
\usepackage{indentfirst}
\usepackage{latexsym}
\usepackage{mathrsfs}
\usepackage{mciteplus}
\usepackage{multirow}
\usepackage{rotating}
\usepackage{scrextend}
\usepackage{soul}
\usepackage{subfigure}
\usepackage{verbatim}

\begin{document}

\title[Darboux Covariance: A Hidden Symmetry of Perturbed Schwarzschild Black Holes]{Darboux Covariance: A Hidden Symmetry of Perturbed Schwarzschild Black Holes}

\author{Michele Lenzi}
\affiliation{Institut de Ci\`encies de l'Espai (ICE, CSIC), Campus UAB, Carrer de Can Magrans s/n, 08193 Cerdanyola del Vall\`es, Spain}
\affiliation{Institut d'Estudis Espacials de Catalunya (IEEC), Edifici Nexus, Carrer del Gran Capit\`a 2-4, despatx 201, 08034 Barcelona, Spain}
\affiliation{Dipartimento di Fisica e Astronomia, Universit\`a di Bologna, via Irnerio 46, 40126 Bologna, Italy}

\author{Carlos F. Sopuerta}
\affiliation{Institut de Ci\`encies de l'Espai (ICE, CSIC), Campus UAB, Carrer de Can Magrans s/n, 08193 Cerdanyola del Vall\`es, Spain}
\affiliation{Institut d'Estudis Espacials de Catalunya (IEEC), Edifici Nexus, Carrer del Gran Capit\`a 2-4, despatx 201, 08034 Barcelona, Spain}

%
\begin{abstract}
Starting from the infinite set of possible master equations for the perturbations of Schwarzschild black holes, with master functions linear in the metric perturbations and their first-order derivatives, we show that of all them are connected via Darboux transformations. These transformations preserve physical quantities like the quasinormal mode frequencies and the infinite hierarchy of Korteweg-de Vries conserved quantities, revealing a new hidden symmetry in the description of the perturbations of Schwarzschild black holes: Darboux covariance.
\end{abstract}

\maketitle

%
\noindent{\bf\em Introduction}. General relativistic perturbation theory of spherically-symmetric spacetimes is of paramount importance since it applies to a wide variety of physical phenomena: From structure formation in the homogeneous and isotropic standard cosmological models~\cite{Bardeen:1980PhRvD..22.1882B,Kodama:1984PThPS..78....1K,Bruni:1989PhRvD..40.1804E,Hu:1994jd} to the dynamics of perturbed Schwarzschild black holes (BHs) and spherical relativistic stars~\cite{Nollert:1999re,Kokkotas:1999bd,Andersson:2000mf,Kokkotas:2003mh} (see~\cite{Barack:2018yly} for the impact on gravitational wave physics). In the case of BHs~\cite{Regge:1957td,Zerilli:1970la,Zerilli:1970se,Chandrasekhar:1975nkd}, perturbation theory describes scattering processes~\cite{Chandrasekhar:1992bo,Futterman:1988ni,Andersson:2000tf} and
quasinormal modes~\cite{Nollert:1999re,Kokkotas:1999bd,Berti:2009kk} (QNMs), which are crucial for the last stage of the emission of gravitational radiation from BH binary coalescence.

~

%
\noindent{\bf\em Master Functions and Equations}. Spherically-symmetric spacetimes have a warped geometry and as such the metric has the form: $g^{(4)} = h \times_r \Omega$, where $h_{ab}$ [$x^a=(t,r)$] is a Lorentzian metric, $r$ is the area radial coordinate, and $\Omega_{AB}$ [$x^A=(\theta,\varphi)$] is the metric of the unit 2-sphere.
For a Schwarzschild BH in Schwarzschild coordinates: $h_{ab}={\rm diag}(-f,1/f)$ with $f=1-2M/r$ and $M$ is the BH mass.  
The warped geometry allows us to decompose the metric perturbations in spherical harmonics in such a way that modes with different harmonic numbers $(\ell,m)$ and different parity (odd/even parity) decouple from each other.
We can find linear combinations of the metric perturbations and their first-order derivatives, the master functions $\Psi^{\rm even/odd}_{\ell m}$, so that the perturbative Einstein equations become wave equations of the form: $(\square^{}_2 -V_\ell/f)\Psi^{\rm even/odd}_{\ell m} = 0$, where $\square_2\Psi =h^{ab} \Psi_{:ab}$ and $V_\ell(r) \equiv V(r)$ is the $\ell$-dependent potential.
Considering the BH exterior, where there is a timelike Killing vector, $t^a = \partial/\partial t$, and using the radial tortoise coordinate, $dx/dr = 1/f(r)$, the master equations become
\begin{equation}
\left( -\partial^2_t + L^{}_V \right) \Psi = \left( -\partial^2_t + 
\partial^2_x   - V \right) \Psi = 0 \,,  \label{master-wave-equation}
\end{equation}
where $L^{}_V = \partial^2_x -V$ is the well-known Schr\"odinger time-independent operator. 
Physical quantities like the gravitational-wave fluxes of energy and momenta can be estimated exclusively from the master functions, which are gauge-invariant. Other non-gauge invariant quantities like the self-force~\cite{Poisson:2011nh,Barack:2018yvs} require the reconstruction of all metric perturbations, which depends on the choice of perturbative gauge.

We recently found~\cite{Lenzi:2021wpc} all possible master equations assuming the master functions are linear combinations (with coefficients depending only on $r$) of the metric perturbations and their first-order derivatives.
There are two branches of possible pairs of potentials/master functions, $\{(V,\Psi)\}$:
(i) {\em The standard branch}.  We call it the standard branch because it contains a single potential for each parity:
The Regge-Wheeler potential $V_{\rm RW}$ for odd-parity perturbations~\cite{Regge:1957td} and the Zerilli potential $V_{\rm Z}$ for even-parity perturbations~\cite{Zerilli:1970la}. 
The most general master function is a linear combination (with constant coefficients) of two master functions:
\begin{equation}
{}^{}_{S}\Psi^{\rm even}_{\rm odd} = \mathcal{C}^{}_1 \Psi^{\rm ZM}_{\rm CPM}  + \mathcal{C}^{}_2\Psi^{\rm NE}_{\rm RW} \,, 
\end{equation}
where $\mathcal{C}^{}_1$ and $\mathcal{C}^{}_2$ are two arbitrary constants. 
In the odd-parity case the two master functions turn out to be the well-known Regge-Wheeler~\cite{Regge:1957td}  and Cunningham-Price-Moncrief master functions~\cite{Cunningham:1978cp,Cunningham:1979px,Cunningham:1980cp}: $(\Psi_{\rm RW},\Psi_{\rm CPM})$. In the even-parity case, the two master functions are the Zerilli-Moncrief master function $\Psi_{\rm ZM}$~\cite{Zerilli:1970se,Zerilli:1970la,Moncrief:1974vm} and another function that to our knowledge was unknown, which we called $\Psi_{\rm NE}$ (new even-parity master function).  All of these master functions are gauge-invariant.  The two master functions in each parity are related by a time derivative: $t^a \Psi^{\rm ZM}_{{\rm CPM}\,,a} = 2\,\Psi^{\rm NE}_{\rm RW}\,$.
(ii) {\em The Darboux branch}.  In this branch, for each parity, there is an infinite set of possible potentials and master functions, $\{(V,\Psi)\}$. The set of possible potentials are determined by a non-linear second-order ordinary differential equation.
Then, for each potential the master function can be written as~\cite{Lenzi:2021wpc} 
\begin{equation}
{}^{}_{D}\Psi^{\rm even}_{\rm odd} = \mathcal{C}^{}_1 \Psi^{\rm ZM}_{\rm CPM}  + \mathcal{C}^{}_2\left( \Sigma^{\rm even}_{\rm odd}\, \Psi^{\rm ZM}_{\rm CPM} + \Phi^{\rm  even}_{\rm odd} \right)\,,
\label{even-odd-master-functions}
\end{equation}
where $\Sigma^{\rm even}_{\rm odd}=\Sigma^{\rm even}_{\rm odd}(x)$ is a function that contains the integral of the potential; $\Phi^{\rm even}_{\rm odd}=\Phi^{\rm even}_{\rm odd}(t,r)$ 
are linear combinations of the metric perturbations and their first-order derivatives, but only the combination with $\Psi^{\rm ZM}_{\rm CPM}$ in Eq.~\eqref{even-odd-master-functions} is a true master function.

\vspace{6mm}

%
\noindent{\bf\em Darboux Covariance}. To understand this landscape of master equations, or pairs $\{(V,\Psi)\}$, let us first consider the standard branch, where we have only the Regge-Wheeler and Zerilli potentials. It was first noted by Chandrasekhar~\cite{Chandrasekhar:1975zza,Chandrasekhar:1979iz} that these potentials, for both Schwarzschild and Reissner-Nordstr\"om BHs, lead to the same transmission and reflection coefficients (see also~\cite{Heading:1977jh,1980RSPSA.369..425C}) as well as the same spectra of QNMs frequencies. However, it has only been recently realized~\cite{Glampedakis:2017rar} (see also~\cite{Yurov:2018ynn})  that this is a consequence of the master equations being related by a Darboux transformation (DT)~\cite{1999physics...8003D,Darboux:89} (see also~\cite{Matveev:1991ms,Ohmiya:1999}). 

Two any master equations, characterized by pairs $(v,\varphi)$ and $(V,\Psi)$, are related by a DT if the two pairs are related by a transformation of the form 
\begin{eqnarray}
\Psi = \varphi^{}_{,x}  + g\,\varphi \,,\quad
V = v + 2\,g^{}_{,x}\
\label{Darboux-transformation}
\end{eqnarray}
where the DT generating function, $g(x)$, must satisfy the following Riccati equation
\begin{eqnarray}
g^{}_{,x} - g^2 + v = {\cal C} \,,
\label{conditions-on-Darboux-tranformation}
\end{eqnarray}
where ${\cal C}$ is an arbitrary constant. We can write the DT generating function as $g  =  (V+v)_{,x}/(2 (V- v))$. Then, the consistency between the expressions for $g(x)$ and $g_{,x}(x)$ is a second-order non-linear equation
\begin{equation}
\left(\frac{\delta V^{}_{,x}}{\delta V} \right)^{}_{,x}
+ 2 \left(\frac{v^{}_{,x}}{\delta V} \right)^{}_{,x} - \delta V = 0 \,, \label{xdarboux}
\end{equation}
where $\delta V = V-v\,$.  This is precisely the equation that any potential in the Darboux branch should satisfy~\cite{Lenzi:2021wpc}, where $v=V^{\rm Z}_{\rm RW}$. Hence, all master equations in the Darboux branch are connected via a DT to the standard branch, with DT generating functions given by:
\begin{eqnarray}
g^{\rm even}_{\rm odd} = \frac{1}{2}\int dx\, \left(V - V^{\rm Z}_{\rm RW} \right) \,,
\label{exp-g-odd-and-even}
\end{eqnarray}
while the two parities in the standard branch are connected by a DT with generating function:
\begin{equation}
g^{\rm odd\rightarrow even} = \frac{1}{2}\int dx\,\left(V^{}_{\rm Z}   - V^{}_{\rm RW} \right) = - g^{\rm even \rightarrow odd} \,.
\label{exp-g-odd-to-even}
\end{equation}
In conclusion, we have an infinite set of master equations linked by DTs, showing the existence of a hidden symmetry in the perturbations of spherically-symmetric BHs: Darboux covariance~\cite{Matveev:1991ms}.

In this work we have adopted a view of the DT that is more general than the original one, as introduced by Crum~\cite{Crum:10.1093/qmath/6.1.121}, based on Sturm-Liouville problems and where the generating function of the DT is constructed from an eigenfunction. Instead, we apply the DT to wave-type equations~\eqref{master-wave-equation} and consider generating functions that only have to satisfy Eq.~\eqref{conditions-on-Darboux-tranformation}.  We can make contact with the Crum approach by working
in the frequency domain and study single-frequency solutions: $\Psi(t,r) = e^{i\omega t}\psi(x;\omega)$, which obey a time-independent Schr\"odinger equation
\begin{equation}
L^{}_V\psi(x;\omega) = -\omega^2\psi(x;\omega)\,.
\label{time-independent-schroedinger}
\end{equation}
Given a solution $\psi_o(x;\omega_o)$ with eigenvalue $-\omega^2_o$, the function $g(x) = -(\ln \psi_o)_{,x}$ generates a DT that transform Eq.~\eqref{time-independent-schroedinger} into another equation of the same form with the same eigenvalue $-\omega^2$, therefore showing the isospectral character of the DT. The Riccati equation~\eqref{conditions-on-Darboux-tranformation} is automatically satisfied with ${\cal C} = -\omega^2_o$, and so is Eq.~\eqref{xdarboux}. The new master function from~\eqref{Darboux-transformation}, say $\phi$, can be written as $\phi = W[\psi,\psi_{o}]/\psi_o$  where $W[\psi,\psi_{o}]$ denotes the Wronskian of $\psi$ and $\psi_o$.
It turns out that the DT generating function between the Regge-Wheeler and Zerilli-Moncrief master equations, Eq.~\eqref{exp-g-odd-to-even}, can be constructed from one of the algebraically-special solutions of the Regge-Wheeler equation~\cite{Chandrasekhar:1984:10.2307/2397739,MaassenvandenBrink:2000iwh} (see~\cite{Glampedakis:2017rar}), namely
\begin{equation}
\psi^{}_o = \frac{\lambda(r)}{2}\mbox{e}^{-i\omega^{}_\ast x} \,, \quad
\omega^{}_\ast = -i\frac{n^{}_\ell(n^{}_\ell+1)}{3M}\,,
\label{algebraically-special-solution-1}
\end{equation}
where $n^{}_\ell = (\ell+2)(\ell-1)/2$ and $\lambda(r) = 2n^{}_\ell + 6M/r$. The generating function itself is: 
\begin{equation}
G(x)= \frac{6M\,f(r)}{\lambda(r)r^2}\,.
\end{equation}
Following~\cite{Heading:1977jh,1980RSPSA.369..425C}, the Regge-Wheeler and Zerilli potentials can be written in terms of $G(x)$ as
\begin{equation}
V^{\rm Z}_{\rm RW} =  \pm G^{}_{,x} + \alpha G + G^2 \,, \quad
\alpha = \frac{1}{6M}\frac{(\ell+2)!}{(\ell-2)!} \,, \label{nice-form-RW-Z}
\end{equation}
which can be seen as a Riccati equation for $G(x)$. This form of the potentials is reminiscent of Supersymmetric Quantum Mechanics (SUSY QM)~\cite{Witten:1981nf,Cooper:1982dm,Cooper:1994eh} where the quantum description of systems with double degeneracy of energy levels is realized. This is related to the fact that the Schr\"odinger equation~\eqref{time-independent-schroedinger}, for two DT-related potentials [Eq.~\eqref{Darboux-transformation}], can be written in the form 
\begin{equation}
(\partial^2_x - V^{}_{\pm})\psi = -\hat{\omega}^2\psi\,, \quad
V^{}_{\pm} = \pm g^{}_{,x} + g^2 \,,
\label{supersymmetric-equations}
\end{equation}
where $g(x)$ plays the role of the SUSY QM {\em superpotential}, $V_{\pm}$ are {\em partner} potentials,
$\hat{\omega}^2 = \omega^2 - {\cal C}$ is the energy eigenvalue, and we can introduce ladder operators ${A} = \partial^{}_x - g\,$ and ${A}^\dag = -\partial^{}_x - g\,$ that factorize the Hamiltonians $H_{-} = A\cdot{A}^\dag$ and $H_{+} = {A}^\dag \cdot{A}\,$ ($H_{\pm} = -\partial^2_x + V_{\pm}$). 
In the standard branch: $g=G+\alpha/2$ and ${\cal C} = -\alpha^2/4\,$.
This factorization is the key ingredient of the intertwining operator method used in~\cite{Anderson:1991kx} (see also~\cite{Dotti:2008ta,Leung:1999fr}) to study properties of the infinite set of possible potentials that includes the Regge-Wheeler and Zerilli ones.

~

%
\noindent{\bf\em DTs and the Korteweg-de Vries Equation}. In the frequency-domain we can establish the connection with inverse scattering theory following the work by Gardner, Green, Miura and Kruskal~\cite{Gardner:1967wc} (see also~\cite{Miura:1968JMP.....9.1202M,Miura:1968JMP.....9.1204M}), where they discovered a way to solve the initial-value problem for the Korteweg-de Vries (KdV) equation~\cite{Korteweg:10.1080/14786449508620739} 
\begin{equation}
V^{}_{,\tau} = 6 V V^{}_{,x} - V^{}_{,xxx} \,.
\label{KdV-equation}
\end{equation}
by identifying $V$ with the potential of the time-independent Schr\"odinger equation~\eqref{time-independent-schroedinger}. We now show how this connection reveals interesting properties of our Darboux-covariant master equations in the frequency domain. 
The spectrum of $L_{V}$ is twofold~\cite{Deift:1979dt,Marchenko:2011vam}: It has a continuous part, the {\em scattering states}, and a discrete part 
made out of a finite number of discrete negative eigenvalues.  In the case of the Schwarzschild BH, the potentials $V^{\rm ZM}_{\rm RW}(x)$ are positive everywhere and decay to zero at both ends ($x\rightarrow\pm\infty$). Therefore, there are no discrete normalizable states.
Let us now deform the Schr\"odinger equation by introducing the KdV time $\tau$ in the following way: $V(x)\rightarrow V(\tau,x)$, $\psi(x)\rightarrow\psi(\tau,x)$, and $\omega\rightarrow\omega(\tau)\,$. If $V(\tau,x)$ follows the KdV flow we can show that
\begin{eqnarray}
[\partial^2_x - (V - \omega^2)]\Xi = - (\omega^{2})^{}_{,\tau}\psi  \,, 
\label{compatibility-conditions-lax-pair}
\end{eqnarray}
where $\Xi(\tau,x)=\psi^{}_{,\tau}+V^{}_{,x}\psi-2(V + 2\omega^2)\psi^{}_{,x}$.
In the hypothetical case of bound states (not our case), and assuming that $\psi$ and $V$ decay sufficiently fast at $x\rightarrow\pm\infty$, one can show, by multiplying by $\psi$ and integrating over $x\in(-\infty,\infty)$, that $(\omega^{2})_{,\tau}=0\,$. For non-normalizable states we can adopt an approach due to Lax~\cite{Lax:1968fm} consisting in the introduction of a pair of operators, $P_V$ and $L_V$ ({\em Lax pair}), defined by 
\begin{eqnarray}
\psi^{}_{,\tau} = P_V\psi = -4\psi^{}_{,xxx} + 6V\psi^{}_{,x} + 3V^{}_{,x}\psi \,,
\label{Eq2-KdV-like}
\end{eqnarray}
and Eq.~\eqref{time-independent-schroedinger} respectively. A remarkable fact about this Lax pair is that the relation between differential operators, $dL_V/d\tau = [P_V,L_V]\,$, yields the KdV equation [Eq.~\eqref{KdV-equation}]. 
Following~\cite{Matveev:1991ms}, one can show that the pair of equations $(L_V+\omega^2,-\partial_\tau+P_V)\psi = 0$ is invariant under a DT provided the DT generating function satisfies Eq.~\eqref{conditions-on-Darboux-tranformation}
and is KdV-deformed according to
\begin{eqnarray}
g^{}_{,\tau} = -g^{}_{,xxx} + 6(V+g^{}_{,x})g^{}_{,x} \,.
\label{KdV-deformmation-g}
\end{eqnarray}
On the other hand, using the KdV-deformation of $\psi$, Eq.~\eqref{Eq2-KdV-like}, we rewrite Eq.~\eqref{compatibility-conditions-lax-pair} in the form
\begin{equation}
( V^{}_{,\tau} - 6VV^{}_{,x} +V^{}_{,xxx} - (\omega^{2})^{}_{,\tau} )\psi = 0 \,.
\label{KdV-and-eigenvalue-deformation}
\end{equation}
Therefore, if $(V,\psi)$ are KdV-deformed according to Eqs.~\eqref{KdV-equation} and~\eqref{Eq2-KdV-like}, we must have $\psi\,(\omega^{2})^{}_{,\tau} = 0$, which means that $\omega$ is preserved by the KdV flow. This argument can be applied to the discrete and continuous spectra as well as to the QNM frequencies.

~

%
\noindent{\bf\em DTs and the KdV Hierarchy}. It was shown by Lax~\cite{Lax:1968fm} that equations that are equivalent to a relation between a Lax pair of operators, like the KdV equation, have an infinite set of first integrals. Gardner showed~\cite{gardner1971korteweg} that these first integrals are associated with symmetries of the KdV equation that yield higher-order KdV equations, and all of them can be formulated as a Hamiltonian system with infinite degrees of freedom. Zakharov and Fadeev~\cite{Zakharov:1971faa} showed that the hierarchy of KdV equations leads to a completely integrable Hamiltonian system that admits canonical action-angle variables constructed from the scattering data of the Schr\"odinger equation. 
Here, we use this point of view to study the KdV hierarchy of first integrals for the infinite set of master equations for BH perturbations.

The scattering states of the continuum spectrum coming either from $x\rightarrow-\infty$ or from $x\rightarrow+\infty$ towards the potential barrier described by $V$ are part reflected and part transmitted. For plane waves coming from $x\rightarrow\infty$, the solution of the Schr\"odinger equation has the {\em Jost} asymptotic behavior~\cite{Newton:1982qc}:
\begin{eqnarray}
\psi~\rightarrow~\left\{ \begin{array}{lc}
e^{i\omega x}  &  \mbox{for~}x\rightarrow -\infty\,, \\[2mm]
a(\tau,\omega) e^{i\omega x} + b(\tau,\omega) e^{-i\omega x}  &  \mbox{for~}x\rightarrow +\infty\,,
\end{array} \right. 
\label{continuum-spectrum}
\end{eqnarray}
where the complex coefficients $a(\tau,\omega)$ and $b(\tau,\omega)$, which fully determine the S-matrix, satisfy: $|a|^2 - |b|^2 = 1\,$. The reflection coefficient is $R(\tau,\omega) = b(\tau,\omega)/a(\tau,\omega)$, and in our case, it completely characterizes the scattering data 
so that the mapping $V(\tau,x) \rightarrow s(\tau)$ is uniquely invertible~\cite{Eckhaus:1981tn}.
The transmission coefficient is $T(\tau,\omega)=1/a(\tau,\omega)$ so that $|R|^2+|T|^2=1\,$.  Under the KdV flow they evolve as~\cite{Gardner:1967wc}: $T^{}_{,\tau} = 0$ and $R^{}_{,\tau} = 8i \omega^3 R\,$, which implies:
\begin{equation}
a^{}_{,\tau}=0\,,\quad \mbox{and}\quad b^{}_{,\tau} = 8i \omega^3b\,. 
\label{conservation-law-a}
\end{equation} 
In the inverse scattering method, given the initial value of the potential $V(\tau=0,x)$ we construct the associated scattering data, $s(0)$, evolve it according the KdV flow, thus obtaining $s(\tau)$, and we recover $V(\tau,x)$ from $s(\tau)$ using the Gelfand-Levitan-Marchenko method~\cite{Gelfand:1951gl,marchenko1955reconstruction,Kay:1956km}.

Let us look at the consequences of the conservation law for $a(\tau,\omega)$ [Eq.~\eqref{conservation-law-a}] under the KdV flow. Following~\cite{Zakharov:1971faa}, let us write
\begin{equation}
\psi(\tau,x,\omega) = \exp\left\{i\omega x + \int^x_{-\infty} \!\! dx' \Phi(\tau,x',\omega) \right\} \,,    \label{phase-for-psi}
\end{equation}
so that $a(\tau,\omega)$ becomes [see Eq.~\eqref{continuum-spectrum}] 
\begin{eqnarray}
a(\tau,\omega) =  \lim_{x\rightarrow \infty} e^{-i\omega x} \psi 
           =   \exp\left\{\int_{-\infty}^{+\infty}\!\!\!\!\! dx' \Phi(\tau,x',\omega) \right\} \,.
\end{eqnarray}
It turns out~\cite{Zakharov:1971faa} that $\ln a(\tau,\omega)$ admits an expansion in inverse powers of $\omega$ for $|\omega|\rightarrow\infty$. We can then write
\begin{equation}
\Phi(\tau,x,\omega) = \sum_{n=1}^\infty \frac{f^{}_n(\tau,x)}{(2i\omega)^n} \,. \label{expansion-phase}
\end{equation}
Therefore, $a_{,\tau}=0$ implies that each coefficient $f_n(\tau,x)$ 
yields a conserved quantity, the KdV integrals~\cite{Miura:1968JMP.....9.1204M}: $I_n(\tau)  = \int^{+\infty}_{-\infty} dx f_n(\tau,x)$ with $dI_n/d\tau = 0\,$. After inserting Eq.~\eqref{phase-for-psi} into the Schr\"odinger equation we get
\begin{equation}
\Phi^{}_{,x} + 2i\omega\Phi + \Phi^2 = V \,. \label{Riccati-equation-complex}
\end{equation}
This is a complex Riccati equation.  Introducing the expansion for $\Phi$ here we find that $f_1(\tau,x) = V(\tau,x)$ and a recursion for the rest of coefficients $f^{}_n(\tau,x)$ ($n>1$), which turn out to be differential polynomials in  $V(\tau,x)$:
\begin{equation}
\frac{df^{}_n}{dx} + f^{}_{n+1} + \sum_{m=1}^{n-1} f^{}_m f^{}_{n-m} = 0 \,.
\end{equation}
It is convenient to split $\Phi$ into its real and imaginary parts 
\begin{eqnarray}
\Phi = \sum_{N=1}^\infty \frac{f^{}_{2N}}{(2i\omega)^{2N}} + \sum_{M=0}^\infty  \frac{f^{}_{2M+1}}{(2i\omega)^{2M+1}} =  \chi^{}_{\rm R} + i\chi^{}_{\rm I} \,. 
\label{real-and-imaginary-parts-phase}
\end{eqnarray}
From Eq.~\eqref{Riccati-equation-complex}, $\chi^{}_{\rm R}(\tau,x)$ and $\chi^{}_{\rm I}(\tau,x)$ satisfy 
\begin{eqnarray}
&& \chi^{}_{{\rm R},x} - 2\omega\chi^{}_{\rm I} + \chi^2_{\rm R} - \chi^2_{\rm I} = V\,, 
\label{expansion-phase-real-part} \\
&& \chi^{}_{{\rm I},x} + 2\omega\chi^{}_{\rm R} + 2\chi^{}_{\rm R} \chi^{}_{\rm I} = 0 \,,
\label{expansion-phase-imaginary-part}
\end{eqnarray}
and from here we get an expression for $\chi_{\rm R}$
\begin{equation}
\chi^{}_{\rm R}  = -\frac{1}{2}\frac{d}{dx}\ln(\chi^{}_{\rm I} + \omega) \,,
\label{real-part-is-a-gradient}
\end{equation}
that is, $\chi_{\rm R}$ is a gradient involving only $\chi_{\rm I}$. This, together with the decaying behaviour of the potential $V$, which follows from the decaying properties of $V_{\rm RW}$ and $V_{\rm Z}$ and the Riccati equation~\eqref{conditions-on-Darboux-tranformation}, implies the known result~\cite{Zakharov:1971faa,Eckhaus:1981tn} that all the even KdV integrals, $I_{2N}$, vanish. 
To study the odd KdV integrals, let us integrate Eq.~\eqref{Riccati-equation-complex} over the real line $x\in(-\infty,+\infty)$ and use the decaying properties of our potentials and derivatives to obtain:
\begin{equation}
2i\omega\int^{+\infty}_{-\infty}dx \Phi + \int^{+\infty}_{-\infty}dx\Phi^2 = \int^{+\infty}_{-\infty}dx\, V \,. \label{holy-grail}
\end{equation}
For the standard branch, the potential $V=V^{\rm Z}_{\rm RW}$ admits the form in Eq.~\eqref{nice-form-RW-Z}. Therefore, using the behaviour of $G(x)$ at $x\rightarrow\pm\infty$, the right-hand side of Eq.~\eqref{holy-grail} becomes 
\begin{equation}
\int^{+\infty}_{-\infty}dx\, V = \int^{+\infty}_{-\infty}dx\, (\alpha G + G^2) \,,
\label{holy-grail-2}
\end{equation}
and hence Eq.~\eqref{holy-grail} is the same for the whole standard branch. Any potential of the Darboux branch can be written as $V=V^{\rm even}_{\rm odd} = V^{\rm Z}_{\rm RW} + 2 g^{}_{,x}$. Then, 
using again the decaying properties of the DT generating functions,
we deduce that any potential of the Darboux branch also satisfies Eq.~\eqref{holy-grail-2}, thus Eq.~\eqref{holy-grail} is universal.
We can write it in terms of $(\chi^{}_{\rm R},\chi^{}_{\rm I})$ and then use Eq.~\eqref{real-part-is-a-gradient} and
\begin{eqnarray}
\chi^{}_{\rm R} \chi^{}_{\rm I} = -\frac{1}{2} 
\partial^{}_x\left(\chi^{}_{\rm I} - \omega\ln(\chi^{}_{\rm I} + \omega) \right) \,,
\end{eqnarray}
which is a total derivative. Then, we arrive at
\begin{equation}
-2\omega\int^{+\infty}_{-\infty}\!\!\!\!\!\!\!\!  dx\, \chi^{}_{\rm I} + \int^{+\infty}_{-\infty}\!\!\!\!\!\!\!  dx (\chi^2_{\rm R} -\chi^2_{\rm I}) = \int^{+\infty}_{-\infty}\!\!\!\!\!\!\!  dx (\alpha G + G^2)  \,.
\end{equation}
When we introduce the expansions for $\chi^{}_{\rm R}$ and $\chi^{}_{\rm I}$ [Eq.~\eqref{real-and-imaginary-parts-phase}] this becomes a universal recurrent relation for the odd KdV integrals. 
Then, we conclude that all the KdV integrals associated with the potentials of the infinite set of master equations are the same. 
The first indication of this result was provided by Chandrasekhar's work~\cite{1980RSPSA.369..425C,Chandrasekhar:1992bo}, where he showed evidence that the KdV integrals should be the same for a pair of potentials of the form in Eq.~\eqref{nice-form-RW-Z} (but not necessary related to BH perturbations), although a full proof was not given.

It is interesting to mention that this infinite set of KdV integrals, which makes the KdV equation completely integrable~\cite{Zakharov:1971faa}, has been connected to a recurrence between the infinite KdV hierarchy of equations, initially suggested by Lenard~\cite{Praught:2005SIGMA...1..005P}, and is rooted in the fact that the KdV equation admits a bi-Hamiltonian structure~\cite{gardner1971korteweg,Magri:1977gn}. On the other hand, Gelfand and Dickii~\cite{Gelfand:1975rn} showed that these conserved quantities are connected with trace formulae for half-integer powers of the operator $L_V$.

~

%
\noindent{\bf\em DTs and QNMs}. We have seen that all the possible potentials associated with the infinite set of master equations found in~\cite{Lenzi:2021wpc} have the same set of KdV integrals when studying the continuous spectrum associated with the Sturm-Liouville problem that emerges when one considers scattering problems. QNMs are not associated with a Sturm-Liouville problem, they rather appear as {\em scattering resonances}~\cite{dyatlov2019mathematical}, poles in the meromorphic continuation of the resolvent/Green function (related to $L_V$ in our case).
They can also be seen as the poles of the S-matrix and the associated residues~\cite{Leaver:1986gd}.
We can use the argument given by Chandrashekar~\cite{Chandrasekhar:1992bo} to show that the frequencies of QNMs are the same for all possible potentials, provided they have similar decaying behaviour at $x\rightarrow\pm\infty$. This is the case for our set of potentials by virtue the Riccati equation~\eqref{conditions-on-Darboux-tranformation} for the DT generating function.  Finally, thanks to Eq.~\eqref{KdV-and-eigenvalue-deformation} we can state that QNM frequencies and damping times are preserved by the KdV flow provided the potential is KdV-deformed and the radial master function $\psi$ is KdV-deformed according to Eq.~\eqref{Eq2-KdV-like}.  
Apart from these results, it would be interesting to explore the structure of the resolvent associated with our time-independent master equations and their behaviour and properties under the KdV flow.

~

%
\noindent{\bf\em Conclusions}.  The general structure of master functions and equations has revealed a hidden symmetry in the theory of perturbations of (spherically symmetric) BHs, Darboux covariance. The implications are diverse and here we have shown that, given the decaying properties of the potentials at both infinities (horizon and spatial infinity), DTs preserve the spectrum of scattering processes and QNMs. 
A large part of the developments shown in this work can be extended to other spherically-symmetric backgrounds and even to other theories of gravity. The main changes may come from  different boundary conditions and their implications for the asymptotic behaviour of the potentials.

~

%
\noindent{\bf\em Acknowledgments}. CFS is supported by contracts ESP2017-90084-P and PID2019-106515GB-I00/AEI/10.13039/501100011033 (Spanish Ministry of Science and Innovation) and 2017-SGR-1469 (AGAUR, Generalitat de Catalunya). 
We thank the COST Action CA16104 Gravitational waves, black holes and fundamental physics (GWverse) for a Short Term Scientific Mission award to ML.
ML also acknowledges support from an “Angelo della Riccia” fellowship.

%

\begin{thebibliography}{62}%
\makeatletter
\providecommand \@ifxundefined [1]{%
 \@ifx{#1\undefined}
}%
\providecommand \@ifnum [1]{%
 \ifnum #1\expandafter \@firstoftwo
 \else \expandafter \@secondoftwo
 \fi
}%
\providecommand \@ifx [1]{%
 \ifx #1\expandafter \@firstoftwo
 \else \expandafter \@secondoftwo
 \fi
}%
\providecommand \natexlab [1]{#1}%
\providecommand \enquote  [1]{``#1''}%
\providecommand \bibnamefont  [1]{#1}%
\providecommand \bibfnamefont [1]{#1}%
\providecommand \citenamefont [1]{#1}%
\providecommand \href@noop [0]{\@secondoftwo}%
\providecommand \href [0]{\begingroup \@sanitize@url \@href}%
\providecommand \@href[1]{\@@startlink{#1}\@@href}%
\providecommand \@@href[1]{\endgroup#1\@@endlink}%
\providecommand \@sanitize@url [0]{\catcode `\\12\catcode `\$12\catcode
  `\&12\catcode `\#12\catcode `\^12\catcode `\_12\catcode `\%12\relax}%
\providecommand \@@startlink[1]{}%
\providecommand \@@endlink[0]{}%
\providecommand \url  [0]{\begingroup\@sanitize@url \@url }%
\providecommand \@url [1]{\endgroup\@href {#1}{\urlprefix }}%
\providecommand \urlprefix  [0]{URL }%
\providecommand \Eprint [0]{\href }%
\providecommand \doibase [0]{https://doi.org/}%
\providecommand \selectlanguage [0]{\@gobble}%
\providecommand \bibinfo  [0]{\@secondoftwo}%
\providecommand \bibfield  [0]{\@secondoftwo}%
\providecommand \translation [1]{[#1]}%
\providecommand \BibitemOpen [0]{}%
\providecommand \bibitemStop [0]{}%
\providecommand \bibitemNoStop [0]{.\EOS\space}%
\providecommand \EOS [0]{\spacefactor3000\relax}%
\providecommand \BibitemShut  [1]{\csname bibitem#1\endcsname}%
\let\auto@bib@innerbib\@empty
\bibitem [{\citenamefont {{Bardeen}}(1980)}]{Bardeen:1980PhRvD..22.1882B}%
  \BibitemOpen
  \bibfield  {author} {\bibinfo {author} {\bibfnamefont {J.~M.}\ \bibnamefont
  {{Bardeen}}},\ }\bibfield  {title} {\bibinfo {title} {{Gauge-invariant
  cosmological perturbations}},\ }\href
  {https://doi.org/10.1103/PhysRevD.22.1882} {\bibfield  {journal} {\bibinfo
  {journal} {Phys. Rev. D}\ }\textbf {\bibinfo {volume} {22}},\ \bibinfo
  {pages} {1882} (\bibinfo {year} {1980})}\BibitemShut {NoStop}%
\bibitem [{\citenamefont {{Kodama}}\ and\ \citenamefont
  {{Sasaki}}(1984)}]{Kodama:1984PThPS..78....1K}%
  \BibitemOpen
  \bibfield  {author} {\bibinfo {author} {\bibfnamefont {H.}~\bibnamefont
  {{Kodama}}}\ and\ \bibinfo {author} {\bibfnamefont {M.}~\bibnamefont
  {{Sasaki}}},\ }\bibfield  {title} {\bibinfo {title} {{Cosmological
  Perturbation Theory}},\ }\href {https://doi.org/10.1143/PTPS.78.1} {\bibfield
   {journal} {\bibinfo  {journal} {Progress of Theoretical Physics Supplement}\
  }\textbf {\bibinfo {volume} {78}},\ \bibinfo {pages} {1} (\bibinfo {year}
  {1984})}\BibitemShut {NoStop}%
\bibitem [{\citenamefont {{Ellis}}\ and\ \citenamefont
  {{Bruni}}(1989)}]{Bruni:1989PhRvD..40.1804E}%
  \BibitemOpen
  \bibfield  {author} {\bibinfo {author} {\bibfnamefont {G.~F.~R.}\
  \bibnamefont {{Ellis}}}\ and\ \bibinfo {author} {\bibfnamefont
  {M.}~\bibnamefont {{Bruni}}},\ }\bibfield  {title} {\bibinfo {title}
  {{Covariant and gauge-invariant approach to cosmological density
  fluctuations}},\ }\href {https://doi.org/10.1103/PhysRevD.40.1804} {\bibfield
   {journal} {\bibinfo  {journal} {Phys. Rev. D}\ }\textbf {\bibinfo {volume}
  {40}},\ \bibinfo {pages} {1804} (\bibinfo {year} {1989})}\BibitemShut
  {NoStop}%
\bibitem [{\citenamefont {Hu}\ and\ \citenamefont
  {Sugiyama}(1995)}]{Hu:1994jd}%
  \BibitemOpen
  \bibfield  {author} {\bibinfo {author} {\bibfnamefont {W.}~\bibnamefont
  {Hu}}\ and\ \bibinfo {author} {\bibfnamefont {N.}~\bibnamefont {Sugiyama}},\
  }\bibfield  {title} {\bibinfo {title} {{Toward understanding CMB anisotropies
  and their implications}},\ }\href {https://doi.org/10.1103/PhysRevD.51.2599}
  {\bibfield  {journal} {\bibinfo  {journal} {Phys. Rev. D}\ }\textbf {\bibinfo
  {volume} {51}},\ \bibinfo {pages} {2599} (\bibinfo {year} {1995})},\ \Eprint
  {https://arxiv.org/abs/astro-ph/9411008} {arXiv:astro-ph/9411008}
  \BibitemShut {NoStop}%
\bibitem [{\citenamefont {{Nollert}}(1999)}]{Nollert:1999re}%
  \BibitemOpen
  \bibfield  {author} {\bibinfo {author} {\bibfnamefont {H.-P.}\ \bibnamefont
  {{Nollert}}},\ }\bibfield  {title} {\bibinfo {title} {{Quasinormal modes: the
  characteristic `sound' of black holes and neutron stars}},\ }\href@noop {}
  {\bibfield  {journal} {\bibinfo  {journal} {Class. Quantum Grav.}\ }\textbf
  {\bibinfo {volume} {16}},\ \bibinfo {pages} {159} (\bibinfo {year}
  {1999})}\BibitemShut {NoStop}%
\bibitem [{\citenamefont {Kokkotas}\ and\ \citenamefont
  {Schmidt}(1999)}]{Kokkotas:1999bd}%
  \BibitemOpen
  \bibfield  {author} {\bibinfo {author} {\bibfnamefont {K.~D.}\ \bibnamefont
  {Kokkotas}}\ and\ \bibinfo {author} {\bibfnamefont {B.~G.}\ \bibnamefont
  {Schmidt}},\ }\bibfield  {title} {\bibinfo {title} {{Quasinormal modes of
  stars and black holes}},\ }\href {https://doi.org/10.12942/lrr-1999-2}
  {\bibfield  {journal} {\bibinfo  {journal} {Living Rev. Rel.}\ }\textbf
  {\bibinfo {volume} {2}},\ \bibinfo {pages} {2} (\bibinfo {year} {1999})},\
  \Eprint {https://arxiv.org/abs/gr-qc/9909058} {arXiv:gr-qc/9909058}
  \BibitemShut {NoStop}%
\bibitem [{\citenamefont {Andersson}\ and\ \citenamefont
  {Kokkotas}(2001)}]{Andersson:2000mf}%
  \BibitemOpen
  \bibfield  {author} {\bibinfo {author} {\bibfnamefont {N.}~\bibnamefont
  {Andersson}}\ and\ \bibinfo {author} {\bibfnamefont {K.~D.}\ \bibnamefont
  {Kokkotas}},\ }\bibfield  {title} {\bibinfo {title} {{The r-mode instability
  in rotating neutron stars}},\ }\href
  {https://doi.org/10.1142/S0218271801001062} {\bibfield  {journal} {\bibinfo
  {journal} {Int. J. Mod. Phys. D}\ }\textbf {\bibinfo {volume} {10}},\
  \bibinfo {pages} {381} (\bibinfo {year} {2001})},\ \Eprint
  {https://arxiv.org/abs/gr-qc/0010102} {arXiv:gr-qc/0010102} \BibitemShut
  {NoStop}%
\bibitem [{\citenamefont {Kokkotas}\ and\ \citenamefont
  {Schutz}(1992)}]{Kokkotas:2003mh}%
  \BibitemOpen
  \bibfield  {author} {\bibinfo {author} {\bibfnamefont {K.~D.}\ \bibnamefont
  {Kokkotas}}\ and\ \bibinfo {author} {\bibfnamefont {B.~F.}\ \bibnamefont
  {Schutz}},\ }\bibfield  {title} {\bibinfo {title} {{W-modes: A New family of
  normal modes of pulsating relativistic stars}},\ }\href@noop {} {\bibfield
  {journal} {\bibinfo  {journal} {Mon. Not. Roy. Astron. Soc.}\ }\textbf
  {\bibinfo {volume} {255}},\ \bibinfo {pages} {119} (\bibinfo {year}
  {1992})}\BibitemShut {NoStop}%
\bibitem [{\citenamefont {Barack}\ \emph {et~al.}(2019)\citenamefont {Barack}
  \emph {et~al.}}]{Barack:2018yly}%
  \BibitemOpen
  \bibfield  {author} {\bibinfo {author} {\bibfnamefont {L.}~\bibnamefont
  {Barack}} \emph {et~al.},\ }\bibfield  {title} {\bibinfo {title} {{Black
  holes, gravitational waves and fundamental physics: a roadmap}},\ }\href
  {https://doi.org/10.1088/1361-6382/ab0587} {\bibfield  {journal} {\bibinfo
  {journal} {Class. Quant. Grav.}\ }\textbf {\bibinfo {volume} {36}},\ \bibinfo
  {pages} {143001} (\bibinfo {year} {2019})},\ \Eprint
  {https://arxiv.org/abs/1806.05195} {arXiv:1806.05195 [gr-qc]} \BibitemShut
  {NoStop}%
\bibitem [{\citenamefont {Regge}\ and\ \citenamefont
  {Wheeler}(1957)}]{Regge:1957td}%
  \BibitemOpen
  \bibfield  {author} {\bibinfo {author} {\bibfnamefont {T.}~\bibnamefont
  {Regge}}\ and\ \bibinfo {author} {\bibfnamefont {J.~A.}\ \bibnamefont
  {Wheeler}},\ }\bibfield  {title} {\bibinfo {title} {{Stability of a
  Schwarzschild singularity}},\ }\href
  {https://doi.org/10.1103/PhysRev.108.1063} {\bibfield  {journal} {\bibinfo
  {journal} {Phys. Rev.}\ }\textbf {\bibinfo {volume} {108}},\ \bibinfo {pages}
  {1063} (\bibinfo {year} {1957})}\BibitemShut {NoStop}%
\bibitem [{\citenamefont {Zerilli}(1970{\natexlab{a}})}]{Zerilli:1970la}%
  \BibitemOpen
  \bibfield  {author} {\bibinfo {author} {\bibfnamefont {F.~J.}\ \bibnamefont
  {Zerilli}},\ }\bibfield  {title} {\bibinfo {title} {{Gravitational Field of a
  Particle Falling in a Schwarzschild Geometry Analyzed in Tensor Harmonics}},\
  }\href@noop {} {\bibfield  {journal} {\bibinfo  {journal} {Phys. Rev. D}\
  }\textbf {\bibinfo {volume} {2}},\ \bibinfo {pages} {2141} (\bibinfo {year}
  {1970}{\natexlab{a}})}\BibitemShut {NoStop}%
\bibitem [{\citenamefont {Zerilli}(1970{\natexlab{b}})}]{Zerilli:1970se}%
  \BibitemOpen
  \bibfield  {author} {\bibinfo {author} {\bibfnamefont {F.~J.}\ \bibnamefont
  {Zerilli}},\ }\bibfield  {title} {\bibinfo {title} {{Effective potential for
  even parity Regge-Wheeler gravitational perturbation equations}},\ }\href
  {https://doi.org/10.1103/PhysRevLett.24.737} {\bibfield  {journal} {\bibinfo
  {journal} {Phys. Rev. Lett.}\ }\textbf {\bibinfo {volume} {24}},\ \bibinfo
  {pages} {737} (\bibinfo {year} {1970}{\natexlab{b}})}\BibitemShut {NoStop}%
\bibitem [{\citenamefont {Chandrasekhar}(1975)}]{Chandrasekhar:1975nkd}%
  \BibitemOpen
  \bibfield  {author} {\bibinfo {author} {\bibfnamefont {S.}~\bibnamefont
  {Chandrasekhar}},\ }\bibfield  {title} {\bibinfo {title} {{On the equations
  governing the perturbations of the Schwarzschild black hole}},\ }\href
  {https://doi.org/10.1098/rspa.1975.0066} {\bibfield  {journal} {\bibinfo
  {journal} {Proc. Roy. Soc. Lond. A}\ }\textbf {\bibinfo {volume} {343}},\
  \bibinfo {pages} {289} (\bibinfo {year} {1975})}\BibitemShut {NoStop}%
\bibitem [{\citenamefont {{Chandrasekhar}}(1992)}]{Chandrasekhar:1992bo}%
  \BibitemOpen
  \bibfield  {author} {\bibinfo {author} {\bibfnamefont {S.}~\bibnamefont
  {{Chandrasekhar}}},\ }\href@noop {} {\emph {\bibinfo {title} {{The
  Mathematical Theory of Black Holes}}}}\ (\bibinfo  {publisher} {Oxford
  University Press},\ \bibinfo {address} {New York},\ \bibinfo {year}
  {1992})\BibitemShut {NoStop}%
\bibitem [{\citenamefont {Futterman}\ \emph {et~al.}(2012)\citenamefont
  {Futterman}, \citenamefont {Handler},\ and\ \citenamefont
  {Matzner}}]{Futterman:1988ni}%
  \BibitemOpen
  \bibfield  {author} {\bibinfo {author} {\bibfnamefont {J.~A.~H.}\
  \bibnamefont {Futterman}}, \bibinfo {author} {\bibfnamefont {F.~A.}\
  \bibnamefont {Handler}},\ and\ \bibinfo {author} {\bibfnamefont {R.~A.}\
  \bibnamefont {Matzner}},\ }\href {https://doi.org/10.1017/CBO9780511735615}
  {\emph {\bibinfo {title} {{Scattering from Black Holes}}}}\ (\bibinfo
  {publisher} {Cambridge University Press},\ \bibinfo {year}
  {2012})\BibitemShut {NoStop}%
\bibitem [{\citenamefont {Andersson}\ and\ \citenamefont
  {Jensen}(2000)}]{Andersson:2000tf}%
  \BibitemOpen
  \bibfield  {author} {\bibinfo {author} {\bibfnamefont {N.}~\bibnamefont
  {Andersson}}\ and\ \bibinfo {author} {\bibfnamefont {B.~P.}\ \bibnamefont
  {Jensen}},\ }\bibfield  {title} {\bibinfo {title} {{Scattering by black
  holes. Chapter 0.1}},\ }\href@noop {} {\  (\bibinfo {year} {2000})},\ \Eprint
  {https://arxiv.org/abs/gr-qc/0011025} {arXiv:gr-qc/0011025} \BibitemShut
  {NoStop}%
\bibitem [{\citenamefont {Berti}\ \emph {et~al.}(2009)\citenamefont {Berti},
  \citenamefont {Cardoso},\ and\ \citenamefont {Starinets}}]{Berti:2009kk}%
  \BibitemOpen
  \bibfield  {author} {\bibinfo {author} {\bibfnamefont {E.}~\bibnamefont
  {Berti}}, \bibinfo {author} {\bibfnamefont {V.}~\bibnamefont {Cardoso}},\
  and\ \bibinfo {author} {\bibfnamefont {A.~O.}\ \bibnamefont {Starinets}},\
  }\bibfield  {title} {\bibinfo {title} {{Quasinormal modes of black holes and
  black branes}},\ }\href {https://doi.org/10.1088/0264-9381/26/16/163001}
  {\bibfield  {journal} {\bibinfo  {journal} {Class. Quant. Grav.}\ }\textbf
  {\bibinfo {volume} {26}},\ \bibinfo {pages} {163001} (\bibinfo {year}
  {2009})},\ \Eprint {https://arxiv.org/abs/0905.2975} {arXiv:0905.2975
  [gr-qc]} \BibitemShut {NoStop}%
\bibitem [{\citenamefont {Poisson}\ \emph {et~al.}(2011)\citenamefont
  {Poisson}, \citenamefont {Pound},\ and\ \citenamefont
  {Vega}}]{Poisson:2011nh}%
  \BibitemOpen
  \bibfield  {author} {\bibinfo {author} {\bibfnamefont {E.}~\bibnamefont
  {Poisson}}, \bibinfo {author} {\bibfnamefont {A.}~\bibnamefont {Pound}},\
  and\ \bibinfo {author} {\bibfnamefont {I.}~\bibnamefont {Vega}},\ }\bibfield
  {title} {\bibinfo {title} {{The Motion of point particles in curved
  spacetime}},\ }\href {https://doi.org/10.12942/lrr-2011-7} {\bibfield
  {journal} {\bibinfo  {journal} {Living Rev. Rel.}\ }\textbf {\bibinfo
  {volume} {14}},\ \bibinfo {pages} {7} (\bibinfo {year} {2011})},\ \Eprint
  {https://arxiv.org/abs/1102.0529} {arXiv:1102.0529 [gr-qc]} \BibitemShut
  {NoStop}%
\bibitem [{\citenamefont {Barack}\ and\ \citenamefont
  {Pound}(2019)}]{Barack:2018yvs}%
  \BibitemOpen
  \bibfield  {author} {\bibinfo {author} {\bibfnamefont {L.}~\bibnamefont
  {Barack}}\ and\ \bibinfo {author} {\bibfnamefont {A.}~\bibnamefont {Pound}},\
  }\bibfield  {title} {\bibinfo {title} {{Self-force and radiation reaction in
  general relativity}},\ }\href {https://doi.org/10.1088/1361-6633/aae552}
  {\bibfield  {journal} {\bibinfo  {journal} {Rept. Prog. Phys.}\ }\textbf
  {\bibinfo {volume} {82}},\ \bibinfo {pages} {016904} (\bibinfo {year}
  {2019})},\ \Eprint {https://arxiv.org/abs/1805.10385} {arXiv:1805.10385
  [gr-qc]} \BibitemShut {NoStop}%
\bibitem [{\citenamefont {Lenzi}\ and\ \citenamefont
  {Sopuerta}(2021)}]{Lenzi:2021wpc}%
  \BibitemOpen
  \bibfield  {author} {\bibinfo {author} {\bibfnamefont {M.}~\bibnamefont
  {Lenzi}}\ and\ \bibinfo {author} {\bibfnamefont {C.~F.}\ \bibnamefont
  {Sopuerta}},\ }\bibfield  {title} {\bibinfo {title} {{Master Functions and
  Equations for Perturbations of Vacuum Spherically-Symmetric Spacetimes}},\
  }\href@noop {} {\  (\bibinfo {year} {2021})},\ \Eprint
  {https://arxiv.org/abs/2108.08668} {arXiv:2108.08668 [gr-qc]} \BibitemShut
  {NoStop}%
\bibitem [{\citenamefont {{Cunningham}}\ \emph {et~al.}(1978)\citenamefont
  {{Cunningham}}, \citenamefont {{Price}},\ and\ \citenamefont
  {{Moncrief}}}]{Cunningham:1978cp}%
  \BibitemOpen
  \bibfield  {author} {\bibinfo {author} {\bibfnamefont {C.~T.}\ \bibnamefont
  {{Cunningham}}}, \bibinfo {author} {\bibfnamefont {R.~H.}\ \bibnamefont
  {{Price}}},\ and\ \bibinfo {author} {\bibfnamefont {V.}~\bibnamefont
  {{Moncrief}}},\ }\bibfield  {title} {\bibinfo {title} {{Radiation from
  collapsing relativistic stars. I - Linearized odd-parity radiation}},\
  }\href@noop {} {\bibfield  {journal} {\bibinfo  {journal} {Astrophys. J.}\
  }\textbf {\bibinfo {volume} {224}},\ \bibinfo {pages} {643} (\bibinfo {year}
  {1978})}\BibitemShut {NoStop}%
\bibitem [{\citenamefont {Cunningham}\ \emph {et~al.}(1979)\citenamefont
  {Cunningham}, \citenamefont {Price},\ and\ \citenamefont
  {Moncrief}}]{Cunningham:1979px}%
  \BibitemOpen
  \bibfield  {author} {\bibinfo {author} {\bibfnamefont {C.~T.}\ \bibnamefont
  {Cunningham}}, \bibinfo {author} {\bibfnamefont {R.~H.}\ \bibnamefont
  {Price}},\ and\ \bibinfo {author} {\bibfnamefont {V.}~\bibnamefont
  {Moncrief}},\ }\bibfield  {title} {\bibinfo {title} {{Radiation from
  collapsing relativistic stars. II. Linearized even parity radiation}},\
  }\href@noop {} {\bibfield  {journal} {\bibinfo  {journal} {Astrophys. J.}\
  }\textbf {\bibinfo {volume} {230}},\ \bibinfo {pages} {870} (\bibinfo {year}
  {1979})}\BibitemShut {NoStop}%
\bibitem [{\citenamefont {{Cunningham}}\ \emph {et~al.}(1980)\citenamefont
  {{Cunningham}}, \citenamefont {{Price}},\ and\ \citenamefont
  {{Moncrief}}}]{Cunningham:1980cp}%
  \BibitemOpen
  \bibfield  {author} {\bibinfo {author} {\bibfnamefont {C.~T.}\ \bibnamefont
  {{Cunningham}}}, \bibinfo {author} {\bibfnamefont {R.~H.}\ \bibnamefont
  {{Price}}},\ and\ \bibinfo {author} {\bibfnamefont {V.}~\bibnamefont
  {{Moncrief}}},\ }\bibfield  {title} {\bibinfo {title} {{Radiation from
  collapsing relativistic stars. III - Second order perturbations of collapse
  with rotation}},\ }\href@noop {} {\bibfield  {journal} {\bibinfo  {journal}
  {Astrophys. J.}\ }\textbf {\bibinfo {volume} {236}},\ \bibinfo {pages} {674}
  (\bibinfo {year} {1980})}\BibitemShut {NoStop}%
\bibitem [{\citenamefont {Moncrief}(1974)}]{Moncrief:1974vm}%
  \BibitemOpen
  \bibfield  {author} {\bibinfo {author} {\bibfnamefont {V.}~\bibnamefont
  {Moncrief}},\ }\bibfield  {title} {\bibinfo {title} {{Gravitational
  perturbations of spherically symmetric systems. I. The exterior problem}},\
  }\href@noop {} {\bibfield  {journal} {\bibinfo  {journal} {Ann. Phys.
  (N.Y.)}\ }\textbf {\bibinfo {volume} {88}},\ \bibinfo {pages} {323} (\bibinfo
  {year} {1974})}\BibitemShut {NoStop}%
\bibitem [{\citenamefont {Chandrasekhar}\ and\ \citenamefont
  {Detweiler}(1975)}]{Chandrasekhar:1975zza}%
  \BibitemOpen
  \bibfield  {author} {\bibinfo {author} {\bibfnamefont {S.}~\bibnamefont
  {Chandrasekhar}}\ and\ \bibinfo {author} {\bibfnamefont {S.~L.}\ \bibnamefont
  {Detweiler}},\ }\bibfield  {title} {\bibinfo {title} {{The quasi-normal modes
  of the Schwarzschild black hole}},\ }\href
  {https://doi.org/10.1098/rspa.1975.0112} {\bibfield  {journal} {\bibinfo
  {journal} {Proc. Roy. Soc. Lond. A}\ }\textbf {\bibinfo {volume} {344}},\
  \bibinfo {pages} {441} (\bibinfo {year} {1975})}\BibitemShut {NoStop}%
\bibitem [{\citenamefont {Chandrasekhar}(1979)}]{Chandrasekhar:1979iz}%
  \BibitemOpen
  \bibfield  {author} {\bibinfo {author} {\bibfnamefont {S.}~\bibnamefont
  {Chandrasekhar}},\ }\bibfield  {title} {\bibinfo {title} {{On the Equations
  Governing the Perturbations of the Reissner-Nordstr\"om Black Hole}},\ }\href
  {https://doi.org/10.1098/rspa.1979.0028} {\bibfield  {journal} {\bibinfo
  {journal} {Proc. Roy. Soc. Lond. A}\ }\textbf {\bibinfo {volume} {365}},\
  \bibinfo {pages} {453} (\bibinfo {year} {1979})}\BibitemShut {NoStop}%
\bibitem [{\citenamefont {Heading}(1977)}]{Heading:1977jh}%
  \BibitemOpen
  \bibfield  {author} {\bibinfo {author} {\bibfnamefont {J.}~\bibnamefont
  {Heading}},\ }\bibfield  {title} {\bibinfo {title} {{Resolution of the
  mystery behind Chandrasekhar{\textquotesingle}s black hole
  transformations}},\ }\href {https://doi.org/10.1088/0305-4470/10/6/011}
  {\bibfield  {journal} {\bibinfo  {journal} {J. Phys. A: Math. Gen.}\ }\textbf
  {\bibinfo {volume} {10}},\ \bibinfo {pages} {885} (\bibinfo {year}
  {1977})}\BibitemShut {NoStop}%
\bibitem [{\citenamefont {{Chandrasekhar}}(1980)}]{1980RSPSA.369..425C}%
  \BibitemOpen
  \bibfield  {author} {\bibinfo {author} {\bibfnamefont {S.}~\bibnamefont
  {{Chandrasekhar}}},\ }\bibfield  {title} {\bibinfo {title} {{On
  One-Dimensional Potential Barriers Having Equal Reflexion and Transmission
  Coefficients}},\ }\href {https://doi.org/10.1098/rspa.1980.0008} {\bibfield
  {journal} {\bibinfo  {journal} {Proc. Roy. Soc. Lond. A}\ }\textbf {\bibinfo
  {volume} {369}},\ \bibinfo {pages} {425} (\bibinfo {year}
  {1980})}\BibitemShut {NoStop}%
\bibitem [{\citenamefont {Glampedakis}\ \emph {et~al.}(2017)\citenamefont
  {Glampedakis}, \citenamefont {Johnson},\ and\ \citenamefont
  {Kennefick}}]{Glampedakis:2017rar}%
  \BibitemOpen
  \bibfield  {author} {\bibinfo {author} {\bibfnamefont {K.}~\bibnamefont
  {Glampedakis}}, \bibinfo {author} {\bibfnamefont {A.~D.}\ \bibnamefont
  {Johnson}},\ and\ \bibinfo {author} {\bibfnamefont {D.}~\bibnamefont
  {Kennefick}},\ }\bibfield  {title} {\bibinfo {title} {{Darboux transformation
  in black hole perturbation theory}},\ }\href
  {https://doi.org/10.1103/PhysRevD.96.024036} {\bibfield  {journal} {\bibinfo
  {journal} {Phys. Rev. D}\ }\textbf {\bibinfo {volume} {96}},\ \bibinfo
  {pages} {024036} (\bibinfo {year} {2017})},\ \Eprint
  {https://arxiv.org/abs/1702.06459} {arXiv:1702.06459 [gr-qc]} \BibitemShut
  {NoStop}%
\bibitem [{\citenamefont {Yurov}\ and\ \citenamefont
  {Yurov}(2019)}]{Yurov:2018ynn}%
  \BibitemOpen
  \bibfield  {author} {\bibinfo {author} {\bibfnamefont {A.~V.}\ \bibnamefont
  {Yurov}}\ and\ \bibinfo {author} {\bibfnamefont {V.~A.}\ \bibnamefont
  {Yurov}},\ }\bibfield  {title} {\bibinfo {title} {{A look at the generalized
  Darboux transformations for the quasinormal spectra in Schwarzschild black
  hole perturbation theory: just how general should it be?}},\ }\href
  {https://doi.org/10.1016/j.physleta.2019.05.024} {\bibfield  {journal}
  {\bibinfo  {journal} {Phys. Lett. A}\ }\textbf {\bibinfo {volume} {383}},\
  \bibinfo {pages} {2571} (\bibinfo {year} {2019})},\ \Eprint
  {https://arxiv.org/abs/1809.10279} {arXiv:1809.10279 [gr-qc]} \BibitemShut
  {NoStop}%
\bibitem [{\citenamefont {{Darboux}}(1882)}]{1999physics...8003D}%
  \BibitemOpen
  \bibfield  {author} {\bibinfo {author} {\bibfnamefont {G.}~\bibnamefont
  {{Darboux}}},\ }\bibfield  {title} {\bibinfo {title} {{On a proposition
  relative to linear equations}},\ }\href@noop {} {\bibfield  {journal}
  {\bibinfo  {journal} {C.R. Acad. Sci. Paris}\ }\textbf {\bibinfo {volume}
  {94}},\ \bibinfo {pages} {1456} (\bibinfo {year} {1882})},\ \Eprint
  {https://arxiv.org/abs/physics/9908003} {arXiv:physics/9908003
  [physics.hist-ph]} \BibitemShut {NoStop}%
\bibitem [{\citenamefont {Darboux}(1889)}]{Darboux:89}%
  \BibitemOpen
  \bibfield  {author} {\bibinfo {author} {\bibfnamefont {G.}~\bibnamefont
  {Darboux}},\ }\href@noop {} {\emph {\bibinfo {title} {{Le\c{c}ons sur la
  th\'eorie g\'en\'erale des surfaces et les application g\'eom\'etriques du
  calcul infinit\'esimal. Deuxi\`eme partie}}}}\ (\bibinfo  {publisher}
  {Gauthier Villars et fils},\ \bibinfo {address} {Paris},\ \bibinfo {year}
  {1889})\BibitemShut {NoStop}%
\bibitem [{\citenamefont {Matveev}\ and\ \citenamefont
  {Salle}(1991)}]{Matveev:1991ms}%
  \BibitemOpen
  \bibfield  {author} {\bibinfo {author} {\bibfnamefont {V.~B.}\ \bibnamefont
  {Matveev}}\ and\ \bibinfo {author} {\bibfnamefont {M.~A.}\ \bibnamefont
  {Salle}},\ }\href@noop {} {\emph {\bibinfo {title} {{Darboux Transformations
  and Solitons}}}}\ (\bibinfo  {publisher} {{Springer-Verlag}},\ \bibinfo
  {address} {New York, Berlin, Heidelberg},\ \bibinfo {year}
  {1991})\BibitemShut {NoStop}%
\bibitem [{\citenamefont {Ohmiya}(1999)}]{Ohmiya:1999}%
  \BibitemOpen
  \bibfield  {author} {\bibinfo {author} {\bibfnamefont {M.}~\bibnamefont
  {Ohmiya}},\ }\bibfield  {title} {\bibinfo {title} {{Spectrum of Darboux
  transformation of differential operator}},\ }\href
  {https://doi.org/ojm/1200788875} {\bibfield  {journal} {\bibinfo  {journal}
  {Osaka Journal of Mathematics}\ }\textbf {\bibinfo {volume} {36}},\ \bibinfo
  {pages} {949 } (\bibinfo {year} {1999})}\BibitemShut {NoStop}%
\bibitem [{\citenamefont {Crum}(1955)}]{Crum:10.1093/qmath/6.1.121}%
  \BibitemOpen
  \bibfield  {author} {\bibinfo {author} {\bibfnamefont {M.~M.}\ \bibnamefont
  {Crum}},\ }\bibfield  {title} {\bibinfo {title} {{Associated Stum-Liouville
  Systems}},\ }\href {https://doi.org/10.1093/qmath/6.1.121} {\bibfield
  {journal} {\bibinfo  {journal} {The Quarterly Journal of Mathematics}\
  }\textbf {\bibinfo {volume} {6}},\ \bibinfo {pages} {121} (\bibinfo {year}
  {1955})}\BibitemShut {NoStop}%
\bibitem [{\citenamefont
  {Chandrasekhar}(1984)}]{Chandrasekhar:1984:10.2307/2397739}%
  \BibitemOpen
  \bibfield  {author} {\bibinfo {author} {\bibfnamefont {S.}~\bibnamefont
  {Chandrasekhar}},\ }\bibfield  {title} {\bibinfo {title} {{On Algebraically
  Special Perturbations of Black Holes}},\ }\href
  {http://www.jstor.org/stable/2397739} {\bibfield  {journal} {\bibinfo
  {journal} {Proc. Roy. Soc. Lond. A}\ }\textbf {\bibinfo {volume} {392}},\
  \bibinfo {pages} {1} (\bibinfo {year} {1984})}\BibitemShut {NoStop}%
\bibitem [{\citenamefont {Maassen van~den
  Brink}(2000)}]{MaassenvandenBrink:2000iwh}%
  \BibitemOpen
  \bibfield  {author} {\bibinfo {author} {\bibfnamefont {A.}~\bibnamefont
  {Maassen van~den Brink}},\ }\bibfield  {title} {\bibinfo {title} {{Analytic
  treatment of black hole gravitational waves at the algebraically special
  frequency}},\ }\href {https://doi.org/10.1103/PhysRevD.62.064009} {\bibfield
  {journal} {\bibinfo  {journal} {Phys. Rev. D}\ }\textbf {\bibinfo {volume}
  {62}},\ \bibinfo {pages} {064009} (\bibinfo {year} {2000})},\ \Eprint
  {https://arxiv.org/abs/gr-qc/0001032} {arXiv:gr-qc/0001032} \BibitemShut
  {NoStop}%
\bibitem [{\citenamefont {Witten}(1981)}]{Witten:1981nf}%
  \BibitemOpen
  \bibfield  {author} {\bibinfo {author} {\bibfnamefont {E.}~\bibnamefont
  {Witten}},\ }\bibfield  {title} {\bibinfo {title} {{Dynamical Breaking of
  Supersymmetry}},\ }\href {https://doi.org/10.1016/0550-3213(81)90006-7}
  {\bibfield  {journal} {\bibinfo  {journal} {Nucl. Phys. B}\ }\textbf
  {\bibinfo {volume} {188}},\ \bibinfo {pages} {513} (\bibinfo {year}
  {1981})}\BibitemShut {NoStop}%
\bibitem [{\citenamefont {Cooper}\ and\ \citenamefont
  {Freedman}(1983)}]{Cooper:1982dm}%
  \BibitemOpen
  \bibfield  {author} {\bibinfo {author} {\bibfnamefont {F.}~\bibnamefont
  {Cooper}}\ and\ \bibinfo {author} {\bibfnamefont {B.}~\bibnamefont
  {Freedman}},\ }\bibfield  {title} {\bibinfo {title} {{Aspects of
  Supersymmetric Quantum Mechanics}},\ }\href
  {https://doi.org/10.1016/0003-4916(83)90034-9} {\bibfield  {journal}
  {\bibinfo  {journal} {Annals Phys.}\ }\textbf {\bibinfo {volume} {146}},\
  \bibinfo {pages} {262} (\bibinfo {year} {1983})}\BibitemShut {NoStop}%
\bibitem [{\citenamefont {Cooper}\ \emph {et~al.}(1995)\citenamefont {Cooper},
  \citenamefont {Khare},\ and\ \citenamefont {Sukhatme}}]{Cooper:1994eh}%
  \BibitemOpen
  \bibfield  {author} {\bibinfo {author} {\bibfnamefont {F.}~\bibnamefont
  {Cooper}}, \bibinfo {author} {\bibfnamefont {A.}~\bibnamefont {Khare}},\ and\
  \bibinfo {author} {\bibfnamefont {U.}~\bibnamefont {Sukhatme}},\ }\bibfield
  {title} {\bibinfo {title} {{Supersymmetry and quantum mechanics}},\ }\href
  {https://doi.org/10.1016/0370-1573(94)00080-M} {\bibfield  {journal}
  {\bibinfo  {journal} {Phys. Rept.}\ }\textbf {\bibinfo {volume} {251}},\
  \bibinfo {pages} {267} (\bibinfo {year} {1995})},\ \Eprint
  {https://arxiv.org/abs/hep-th/9405029} {arXiv:hep-th/9405029} \BibitemShut
  {NoStop}%
\bibitem [{\citenamefont {Anderson}\ and\ \citenamefont
  {Price}(1991)}]{Anderson:1991kx}%
  \BibitemOpen
  \bibfield  {author} {\bibinfo {author} {\bibfnamefont {A.}~\bibnamefont
  {Anderson}}\ and\ \bibinfo {author} {\bibfnamefont {R.~H.}\ \bibnamefont
  {Price}},\ }\bibfield  {title} {\bibinfo {title} {{Intertwining of the
  equations of black hole perturbations}},\ }\href
  {https://doi.org/10.1103/PhysRevD.43.3147} {\bibfield  {journal} {\bibinfo
  {journal} {Phys. Rev. D}\ }\textbf {\bibinfo {volume} {43}},\ \bibinfo
  {pages} {3147} (\bibinfo {year} {1991})}\BibitemShut {NoStop}%
\bibitem [{\citenamefont {Dotti}\ and\ \citenamefont
  {Gleiser}(2009)}]{Dotti:2008ta}%
  \BibitemOpen
  \bibfield  {author} {\bibinfo {author} {\bibfnamefont {G.}~\bibnamefont
  {Dotti}}\ and\ \bibinfo {author} {\bibfnamefont {R.~J.}\ \bibnamefont
  {Gleiser}},\ }\bibfield  {title} {\bibinfo {title} {{The Initial value
  problem for linearized gravitational perturbations of the Schwarzchild naked
  singularity}},\ }\href {https://doi.org/10.1088/0264-9381/26/21/215002}
  {\bibfield  {journal} {\bibinfo  {journal} {Class. Quant. Grav.}\ }\textbf
  {\bibinfo {volume} {26}},\ \bibinfo {pages} {215002} (\bibinfo {year}
  {2009})},\ \Eprint {https://arxiv.org/abs/0809.3615} {arXiv:0809.3615
  [gr-qc]} \BibitemShut {NoStop}%
\bibitem [{\citenamefont {Leung}\ \emph {et~al.}(1999)\citenamefont {Leung},
  \citenamefont {Maassen van~den Brink}, \citenamefont {Suen}, \citenamefont
  {Wong},\ and\ \citenamefont {Young}}]{Leung:1999fr}%
  \BibitemOpen
  \bibfield  {author} {\bibinfo {author} {\bibfnamefont {P.~T.}\ \bibnamefont
  {Leung}}, \bibinfo {author} {\bibfnamefont {A.}~\bibnamefont {Maassen van~den
  Brink}}, \bibinfo {author} {\bibfnamefont {W.~M.}\ \bibnamefont {Suen}},
  \bibinfo {author} {\bibfnamefont {C.~W.}\ \bibnamefont {Wong}},\ and\
  \bibinfo {author} {\bibfnamefont {K.}~\bibnamefont {Young}},\ }\bibfield
  {title} {\bibinfo {title} {{SUSY transformations for quasinormal and total
  transmission modes of open systems}},\ }\href@noop {} {\  (\bibinfo {year}
  {1999})},\ \Eprint {https://arxiv.org/abs/math-ph/9909030}
  {arXiv:math-ph/9909030} \BibitemShut {NoStop}%
\bibitem [{\citenamefont {Gardner}\ \emph {et~al.}(1967)\citenamefont
  {Gardner}, \citenamefont {Greene}, \citenamefont {Kruskal},\ and\
  \citenamefont {Miura}}]{Gardner:1967wc}%
  \BibitemOpen
  \bibfield  {author} {\bibinfo {author} {\bibfnamefont {C.~S.}\ \bibnamefont
  {Gardner}}, \bibinfo {author} {\bibfnamefont {J.~M.}\ \bibnamefont {Greene}},
  \bibinfo {author} {\bibfnamefont {M.~D.}\ \bibnamefont {Kruskal}},\ and\
  \bibinfo {author} {\bibfnamefont {R.~M.}\ \bibnamefont {Miura}},\ }\bibfield
  {title} {\bibinfo {title} {{Method for solving the Korteweg-deVries
  equation}},\ }\href {https://doi.org/10.1103/PhysRevLett.19.1095} {\bibfield
  {journal} {\bibinfo  {journal} {Phys. Rev. Lett.}\ }\textbf {\bibinfo
  {volume} {19}},\ \bibinfo {pages} {1095} (\bibinfo {year}
  {1967})}\BibitemShut {NoStop}%
\bibitem [{\citenamefont {{Miura}}(1968)}]{Miura:1968JMP.....9.1202M}%
  \BibitemOpen
  \bibfield  {author} {\bibinfo {author} {\bibfnamefont {R.~M.}\ \bibnamefont
  {{Miura}}},\ }\bibfield  {title} {\bibinfo {title} {{Korteweg-de Vries
  Equation and Generalizations. I. A Remarkable Explicit Nonlinear
  Transformation}},\ }\href {https://doi.org/10.1063/1.1664700} {\bibfield
  {journal} {\bibinfo  {journal} {J. Math. Phys.}\ }\textbf {\bibinfo {volume}
  {9}},\ \bibinfo {pages} {1202} (\bibinfo {year} {1968})}\BibitemShut
  {NoStop}%
\bibitem [{\citenamefont {{Miura}}\ \emph {et~al.}(1968)\citenamefont
  {{Miura}}, \citenamefont {{Gardner}},\ and\ \citenamefont
  {{Kruskal}}}]{Miura:1968JMP.....9.1204M}%
  \BibitemOpen
  \bibfield  {author} {\bibinfo {author} {\bibfnamefont {R.~M.}\ \bibnamefont
  {{Miura}}}, \bibinfo {author} {\bibfnamefont {C.~S.}\ \bibnamefont
  {{Gardner}}},\ and\ \bibinfo {author} {\bibfnamefont {M.~D.}\ \bibnamefont
  {{Kruskal}}},\ }\bibfield  {title} {\bibinfo {title} {{Korteweg-de Vries
  Equation and Generalizations. II. Existence of Conservation Laws and
  Constants of Motion}},\ }\href {https://doi.org/10.1063/1.1664701} {\bibfield
   {journal} {\bibinfo  {journal} {J. Math. Phys.}\ }\textbf {\bibinfo {volume}
  {9}},\ \bibinfo {pages} {1204} (\bibinfo {year} {1968})}\BibitemShut
  {NoStop}%
\bibitem [{\citenamefont {Korteweg}\ and\ \citenamefont
  {de~Vries}(1895)}]{Korteweg:10.1080/14786449508620739}%
  \BibitemOpen
  \bibfield  {author} {\bibinfo {author} {\bibfnamefont {D.~J.}\ \bibnamefont
  {Korteweg}}\ and\ \bibinfo {author} {\bibfnamefont {G.}~\bibnamefont
  {de~Vries}},\ }\bibfield  {title} {\bibinfo {title} {{On the change of form
  of long waves advancing in a rectangular canal, and on a new type of long
  stationary waves}},\ }\href {https://doi.org/10.1080/14786449508620739}
  {\bibfield  {journal} {\bibinfo  {journal} {Philos. Mag.}\ }\textbf {\bibinfo
  {volume} {39}},\ \bibinfo {pages} {422} (\bibinfo {year} {1895})}\BibitemShut
  {NoStop}%
\bibitem [{\citenamefont {Deift}\ and\ \citenamefont
  {Trubowitz}(1979)}]{Deift:1979dt}%
  \BibitemOpen
  \bibfield  {author} {\bibinfo {author} {\bibfnamefont {P.}~\bibnamefont
  {Deift}}\ and\ \bibinfo {author} {\bibfnamefont {E.}~\bibnamefont
  {Trubowitz}},\ }\bibfield  {title} {\bibinfo {title} {{Inverse scattering on
  the line}},\ }\href {https://doi.org/https://doi.org/10.1002/cpa.3160320202}
  {\bibfield  {journal} {\bibinfo  {journal} {Commun. Pure Appl. Math.}\
  }\textbf {\bibinfo {volume} {32}},\ \bibinfo {pages} {121} (\bibinfo {year}
  {1979})}\BibitemShut {NoStop}%
\bibitem [{\citenamefont {Marchenko}(2011)}]{Marchenko:2011vam}%
  \BibitemOpen
  \bibfield  {author} {\bibinfo {author} {\bibfnamefont {V.~A.}\ \bibnamefont
  {Marchenko}},\ }\href@noop {} {\emph {\bibinfo {title} {{Sturm-Liouville
  Operators and Applications}}}}\ (\bibinfo  {publisher} {{AMS Chelsea
  Publishing}},\ \bibinfo {address} {Providence, Rhode Island},\ \bibinfo
  {year} {2011})\BibitemShut {NoStop}%
\bibitem [{\citenamefont {Lax}(1968)}]{Lax:1968fm}%
  \BibitemOpen
  \bibfield  {author} {\bibinfo {author} {\bibfnamefont {P.~D.}\ \bibnamefont
  {Lax}},\ }\bibfield  {title} {\bibinfo {title} {{Integrals of Nonlinear
  Equations of Evolution and Solitary Waves}},\ }\href@noop {} {\bibfield
  {journal} {\bibinfo  {journal} {Commun. Pure Appl. Math.}\ }\textbf {\bibinfo
  {volume} {21}},\ \bibinfo {pages} {467} (\bibinfo {year} {1968})}\BibitemShut
  {NoStop}%
\bibitem [{\citenamefont {Gardner}(1971)}]{gardner1971korteweg}%
  \BibitemOpen
  \bibfield  {author} {\bibinfo {author} {\bibfnamefont {C.~S.}\ \bibnamefont
  {Gardner}},\ }\bibfield  {title} {\bibinfo {title} {{Korteweg-de Vries
  equation and generalizations. IV. The Korteweg-de Vries equation as a
  Hamiltonian system}},\ }\href@noop {} {\bibfield  {journal} {\bibinfo
  {journal} {J. Math. Phys.}\ }\textbf {\bibinfo {volume} {12}},\ \bibinfo
  {pages} {1548} (\bibinfo {year} {1971})}\BibitemShut {NoStop}%
\bibitem [{\citenamefont {Zakharov}\ and\ \citenamefont
  {Faddeev}(1971)}]{Zakharov:1971faa}%
  \BibitemOpen
  \bibfield  {author} {\bibinfo {author} {\bibfnamefont {V.~E.}\ \bibnamefont
  {Zakharov}}\ and\ \bibinfo {author} {\bibfnamefont {L.~D.}\ \bibnamefont
  {Faddeev}},\ }\bibfield  {title} {\bibinfo {title} {{Korteweg-de Vries
  equation: A completely integrable Hamiltonian system}},\ }\href
  {https://doi.org/10.1007/BF01086739} {\bibfield  {journal} {\bibinfo
  {journal} {Functional Analysis and Its Applications}\ }\textbf {\bibinfo
  {volume} {5}},\ \bibinfo {pages} {280} (\bibinfo {year} {1971})}\BibitemShut
  {NoStop}%
\bibitem [{\citenamefont {Newton}(1982)}]{Newton:1982qc}%
  \BibitemOpen
  \bibfield  {author} {\bibinfo {author} {\bibfnamefont {R.~G.}\ \bibnamefont
  {Newton}},\ }\href@noop {} {\emph {\bibinfo {title} {{Scattering Theory of
  Waves and Particles}}}}\ (\bibinfo  {publisher} {Springer Science},\ \bibinfo
  {address} {New York},\ \bibinfo {year} {1982})\BibitemShut {NoStop}%
\bibitem [{\citenamefont {Eckhaus}\ and\ \citenamefont
  {Van~Harten}(1981)}]{Eckhaus:1981tn}%
  \BibitemOpen
  \bibfield  {author} {\bibinfo {author} {\bibfnamefont {W.}~\bibnamefont
  {Eckhaus}}\ and\ \bibinfo {author} {\bibfnamefont {A.}~\bibnamefont
  {Van~Harten}},\ }\href@noop {} {\emph {\bibinfo {title} {{The Inverse
  Scattering Transformation and the Theory of Solitons. An Introduction}}}},\
  \bibinfo {series} {{North-Holland Mathematical Studies}}, Vol.~\bibinfo
  {volume} {50}\ (\bibinfo  {publisher} {{North Holland}},\ \bibinfo {year}
  {1981})\BibitemShut {NoStop}%
\bibitem [{\citenamefont {Gelfand}\ and\ \citenamefont
  {Levitan}(1951)}]{Gelfand:1951gl}%
  \BibitemOpen
  \bibfield  {author} {\bibinfo {author} {\bibfnamefont {I.~M.}\ \bibnamefont
  {Gelfand}}\ and\ \bibinfo {author} {\bibfnamefont {B.~M.}\ \bibnamefont
  {Levitan}},\ }\bibfield  {title} {\bibinfo {title} {{On the determination of
  a differential equation from its spectral function}},\ }\href@noop {}
  {\bibfield  {journal} {\bibinfo  {journal} {Izv. Akad. Nauk SSSR, Ser. Mat.}\
  }\textbf {\bibinfo {volume} {15}},\ \bibinfo {pages} {309} (\bibinfo {year}
  {1951})}\BibitemShut {NoStop}%
\bibitem [{\citenamefont {Marchenko}(1955)}]{marchenko1955reconstruction}%
  \BibitemOpen
  \bibfield  {author} {\bibinfo {author} {\bibfnamefont {V.~A.}\ \bibnamefont
  {Marchenko}},\ }\bibfield  {title} {\bibinfo {title} {{On reconstruction of
  the potential energy from phases of the scattered waves}},\ }in\ \href@noop
  {} {\emph {\bibinfo {booktitle} {Dokl. Akad. Nauk SSSR}}},\ Vol.\ \bibinfo
  {volume} {104}\ (\bibinfo {year} {1955})\ pp.\ \bibinfo {pages}
  {695--698}\BibitemShut {NoStop}%
\bibitem [{\citenamefont {Kay}\ and\ \citenamefont {Moses}(1956)}]{Kay:1956km}%
  \BibitemOpen
  \bibfield  {author} {\bibinfo {author} {\bibfnamefont {I.}~\bibnamefont
  {Kay}}\ and\ \bibinfo {author} {\bibfnamefont {H.}~\bibnamefont {Moses}},\
  }\bibfield  {title} {\bibinfo {title} {{The determination of the scattering
  potential from the spectral measure function}},\ }\href
  {https://doi.org/10.1007/BF02745417} {\bibfield  {journal} {\bibinfo
  {journal} {Nuovo Cim.}\ }\textbf {\bibinfo {volume} {3}},\ \bibinfo {pages}
  {276–304} (\bibinfo {year} {1956})}\BibitemShut {NoStop}%
\bibitem [{\citenamefont {{Praught}}\ and\ \citenamefont
  {{Smirnov}}(2005)}]{Praught:2005SIGMA...1..005P}%
  \BibitemOpen
  \bibfield  {author} {\bibinfo {author} {\bibfnamefont {J.}~\bibnamefont
  {{Praught}}}\ and\ \bibinfo {author} {\bibfnamefont {R.~G.}\ \bibnamefont
  {{Smirnov}}},\ }\bibfield  {title} {\bibinfo {title} {{Andrew Lenard: A
  Mystery Unraveled}},\ }\href {https://doi.org/10.3842/SIGMA.2005.005}
  {\bibfield  {journal} {\bibinfo  {journal} {SIGMA}\ }\textbf {\bibinfo
  {volume} {1}},\ \bibinfo {eid} {005} (\bibinfo {year} {2005})},\ \Eprint
  {https://arxiv.org/abs/nlin/0510055} {arXiv:nlin/0510055 [nlin.SI]}
  \BibitemShut {NoStop}%
\bibitem [{\citenamefont {Magri}(1978)}]{Magri:1977gn}%
  \BibitemOpen
  \bibfield  {author} {\bibinfo {author} {\bibfnamefont {F.}~\bibnamefont
  {Magri}},\ }\bibfield  {title} {\bibinfo {title} {{A Simple model of the
  integrable Hamiltonian equation}},\ }\href {https://doi.org/10.1063/1.523777}
  {\bibfield  {journal} {\bibinfo  {journal} {J. Math. Phys.}\ }\textbf
  {\bibinfo {volume} {19}},\ \bibinfo {pages} {1156} (\bibinfo {year}
  {1978})}\BibitemShut {NoStop}%
\bibitem [{\citenamefont {Gelfand}\ and\ \citenamefont
  {Dikii}(1975)}]{Gelfand:1975rn}%
  \BibitemOpen
  \bibfield  {author} {\bibinfo {author} {\bibfnamefont {I.~M.}\ \bibnamefont
  {Gelfand}}\ and\ \bibinfo {author} {\bibfnamefont {L.~A.}\ \bibnamefont
  {Dikii}},\ }\bibfield  {title} {\bibinfo {title} {{Asymptotic behavior of the
  resolvent of Sturm-Liouville equations and the algebra of the Korteweg-De
  Vries equations}},\ }\href {https://doi.org/10.1070/RM1975v030n05ABEH001522}
  {\bibfield  {journal} {\bibinfo  {journal} {Russ. Math. Surveys}\ }\textbf
  {\bibinfo {volume} {30}},\ \bibinfo {pages} {77} (\bibinfo {year}
  {1975})}\BibitemShut {NoStop}%
\bibitem [{\citenamefont {Dyatlov}\ and\ \citenamefont
  {Zworski}(2019)}]{dyatlov2019mathematical}%
  \BibitemOpen
  \bibfield  {author} {\bibinfo {author} {\bibfnamefont {S.}~\bibnamefont
  {Dyatlov}}\ and\ \bibinfo {author} {\bibfnamefont {M.}~\bibnamefont
  {Zworski}},\ }\href@noop {} {\emph {\bibinfo {title} {Mathematical theory of
  scattering resonances}}},\ Vol.\ \bibinfo {volume} {200}\ (\bibinfo
  {publisher} {American Mathematical Soc.},\ \bibinfo {year}
  {2019})\BibitemShut {NoStop}%
\bibitem [{\citenamefont {Leaver}(1986)}]{Leaver:1986gd}%
  \BibitemOpen
  \bibfield  {author} {\bibinfo {author} {\bibfnamefont {E.~W.}\ \bibnamefont
  {Leaver}},\ }\bibfield  {title} {\bibinfo {title} {{Spectral decomposition of
  the perturbation response of the Schwarzschild geometry}},\ }\href
  {https://doi.org/10.1103/PhysRevD.34.384} {\bibfield  {journal} {\bibinfo
  {journal} {Phys. Rev. D}\ }\textbf {\bibinfo {volume} {34}},\ \bibinfo
  {pages} {384} (\bibinfo {year} {1986})}\BibitemShut {NoStop}%
\end{thebibliography}

%

\end{document}